\documentstyle[prl,aps,multicol]{revtex}
\draft

\author{M.~A.~Nielsen\thanks{mnielsen@theory.caltech.edu}}
\address{Department of Physics, MC 12-33, California Institute of
Technology, Pasadena, CA 91125}
\title{Conditions for a class of entanglement transformations}
\date{\today}

\begin{document}

\pagestyle{plain}
\pagenumbering{arabic}

\maketitle

\begin{abstract}
  Suppose Alice and Bob jointly possess a pure state, $|\psi\rangle$.
  Using local operations on their respective systems and classical
  communication it may be possible for Alice and Bob to {\em
  transform} $|\psi\rangle$ into another joint state $|\phi\rangle$.
  This Letter gives necessary and sufficient conditions for this
  process of {\em entanglement transformation} to be possible.  These
  conditions reveal a partial ordering on the entangled states, and
  connect quantum entanglement to the algebraic theory of {\em
  majorization}.  As a consequence, we find that there exist
  essentially different types of entanglement for bipartite quantum
  systems.
\end{abstract}


\begin{multicols}{2}[]
\narrowtext

%
%
The question ``{\em What tasks may be accomplished using a given
physical resource?}'' is of fundamental importance in many areas of
physics.  In particular, the burgeoning field of quantum information
\cite{Bennett95a,Preskill98a} is much concerned with understanding
transformations between different types of quantum information.  A
fundamental example is the problem of {\em entanglement
transformation}: Suppose $|\psi\rangle$ is a pure state of some composite
system $AB$; we refer to system $A$ as Alice's system, and to system
$B$ as Bob's system.  Into what class of states $|\phi\rangle$ may
$|\psi\rangle$ be {\em transformed}, assuming that Alice and Bob may only
use local operations on their respective systems, and unlimited
two-way classical communication?

%
%
This Letter presents necessary and sufficient conditions for
entanglement transformation to be possible.  These conditions exhibit
an unexpected connection between entanglement and the linear-algebraic
theory of {\em majorization}.  Furthermore, the existence of
essentially different types of entanglement follows immediately from
the conditions, together with a classification for the different
types.

%
%
There has been extensive work on entanglement transformation.  The
problem was introduced in two guises by Bennett {\em et al}
\cite{Bennett96c,Bennett96b,Bennett96a}.  They studied {\em entanglement
distillation}, solving the problem of transforming some given pure
state into (approximate) EPR pairs in the asymptotic limit where many
identical copies of the pure state are initially available.  They also
studied the inverse procedure of {\em entanglement formation}, solving
the problem of transforming EPR pairs into many (approximate) copies
of some given pure state, again in the asymptotic limit.  We will
rederive these results as a consequence of the present investigation.
In \cite{Bennett96c} the problem was also generalized to asymptotic
and approximate transformation between mixed states and EPR pairs, and
further results about these transformations were obtained in
\cite{Bennett96b,Bennett96a}.

%
%
The investigations here are for the finite (non-asymptotic) case, from
which asymptotic results may be recovered by taking limits.  We do not
consider approximate transformations. 

%
%
{\em Majorization} is a large and active area of research in linear
algebra, with entire books \cite{Marshall79a,Alberti82a} devoted to
its theory and application.  We shall use Chapter~2 of Bhatia
\cite{Bhatia97a} as our principal reference on majorization.  Suppose
$x \equiv (x_1,\ldots,x_d)$ and $y = (y_1,\ldots,y_d)$ are real
$d$-dimensional vectors.  Then {\em $x$ is majorized by $y$}
(equivalently {\em $y$ majorizes $x$}), written $x \prec y$, if for
each $k$ in the range $1,\ldots,d$,
\begin{eqnarray}
\sum_{j=1}^k x^{\downarrow}_j \leq \sum_{j=1}^k y^{\downarrow}_j, 
\end{eqnarray} 
with equality holding when $k = d$, and where the $\downarrow$
indicates that elements are to be taken in descending order, so for
example, $x^{\downarrow}_1$ is the largest element in
$(x_1,\ldots,x_d)$.  The majorization relation is a partial order on
real vectors, with $x \prec y$ and $y \prec x$ if and only if
$x^{\downarrow} = y^{\downarrow}$.

%
%
To state our central result linking entanglement with majorization we
need some notation.  Suppose $|\psi\rangle$ is any state of Alice and
Bob's system.  $\rho_\psi$ denotes the state of Alice's system, that
is, $\rho_\psi \equiv \mbox{tr}_B(|\psi\rangle \langle
\psi|)$.  $\lambda_\psi$ denotes the vector of eigenvalues of
$\rho_\psi$.  $|\psi\rangle \rightarrow |\phi\rangle$, read ``$|\psi\rangle$
transforms to $|\phi\rangle$'' indicates that $|\psi\rangle$ may be
transformed into $|\phi\rangle$ by local operations and potentially
unlimited two-way classical communication.  Then we have:

{\bf Theorem 1:} {\em $|\psi\rangle$ transforms to $|\phi\rangle$ using local
operations and classical communication if and only if $\lambda_\psi$
is majorized by $\lambda_\phi$.  More succinctly:
\begin{eqnarray}
|\psi\rangle \rightarrow |\phi\rangle \mbox{ iff } \lambda_{\psi} \prec
\lambda_{\phi}.
\end{eqnarray}
}  

As a simple application of the result, suppose Alice and Bob each
possess a three dimensional quantum system, with respective
orthonormal bases denoted $|1\rangle,|2\rangle,|3\rangle$.  Define states
$|\psi\rangle$ and $|\phi\rangle$ of their joint system by:
\begin{eqnarray} 
|\psi\rangle & \equiv & \sqrt{\frac 12}|11\rangle + \sqrt{\frac 25} |22\rangle +
\sqrt{\frac{1}{10}}|33\rangle \\
|\phi\rangle & \equiv & \sqrt{\frac
  35}|11\rangle + \sqrt{\frac 15} |22\rangle + \sqrt{\frac{1}{5}} |33\rangle.
\end{eqnarray} 
It follows from Theorem~1 that neither $|\psi\rangle \rightarrow
|\phi\rangle$ nor $|\phi\rangle \rightarrow |\psi\rangle$, providing
an example of essentially different types of entanglement, from the
point of view of local operations and classical communication.  We
will say that $|\psi\rangle$ and $|\phi\rangle$ are {\em
incomparable}.  Bennett, Popescu, Rohrlich and Smolin
\cite{Bennett99a} have found specific examples of {\em three-party}
entangled states which are incomparable in a similar sense.

%
%
To prove the theorem we first collect some useful facts:
\begin{enumerate}

\item \label{fact:Lo_Popescu} Lo and Popescu \cite{Lo97b} have
shown that an arbitrary protocol transforming $|\psi\rangle$ to $|\phi\rangle$
using local operations and two-way classical communication may be
simulated by a one-way communication protocol of the following form:
Alice performs a generalized measurement on her system, and then sends
the result of her measurement to Bob, who performs an operation on his
system, conditional on the measurement result.  

\item \label{fact:polar_decomposition} For any matrix $A$, the {\em
polar decomposition} \cite{Bhatia97a} states that $A =
\sqrt{AA^{\dagger}} U$, for some unitary $U$.

\item \label{fact:random_unitary} Suppose $\rho' = \sum_i p_i U_i \rho
U_i^{\dagger}$, where $p_i \geq 0, \sum_i p_i = 1$, and the $U_i$ are
unitary.  Then the vector of eigenvalues of $\rho'$ is majorized by
the vector of eigenvalues of $\rho$, $\lambda_{\rho'} \prec
\lambda_{\rho}$, in an obvious notation \cite{Uhlmann71a}.

\item \label{fact:t-transforms}  Suppose $x \prec y$.  Then $x = Dy$,
where $D$ is a matrix that may be written as a product of at most
$d-1$ {\em T-transforms}, where $d$ is the dimension of $x$ and $y$
\cite{Bhatia97a}.  A T-transform, by definition, acts as the identity
on all but $2$ matrix components.  On those two components it has the
form:
\begin{eqnarray}
T = \left[ \begin{array}{cc} t & 1-t \\ 1-t & t \end{array} \right],
\end{eqnarray}
where $0 \leq t \leq 1$. 

\item \label{fact:Schmidt} 
We make repeated use of the {\em Schmidt decomposition}
\cite{Peres93a}: Any pure state $|x\rangle$ of a
composite system $AB$ may be written in the form $|x\rangle = \sum_i
\sqrt{\lambda_i} |i_A\rangle|i_B\rangle$, where $0 \leq \lambda_i$, $\sum_i
\lambda_i = 1$, and $|i_A\rangle$ ($|i_B\rangle$) form an orthonormal basis
for system $A$ ($B$).  Note that $\rho_x$ has eigenvalues
$\lambda_i$.  Furthermore, we write $|x\rangle \sim |y\rangle$ if $|x\rangle$ and
$|y\rangle$ are the same up to local unitary operations by Alice and Bob.
The Schmidt decomposition implies that $|x\rangle \sim |y\rangle$ if and only
if $\rho_x$ and $\rho_y$ have the same spectrum of eigenvalues.

\end{enumerate}

{\bf Proof of Theorem~1:}

Suppose first that $|\psi\rangle \rightarrow |\phi\rangle$.  Using
fact~\ref{fact:Lo_Popescu} we assume that Alice performs a generalized
measurement \cite{Gardiner91a}, described by operators $M_m$ on her
system, satisfying the {\em completeness relation} $\sum_m
M_m^{\dagger} M_m = I$, and then sends the result to Bob, who performs
an operation ${\cal E}^B_m$, possibly non-unitary, on his system,
conditional on the result $m$.  Thus
\begin{eqnarray}
|\phi\rangle \langle \phi| = \sum_m {\cal E}^B_m \left( M_m |\psi\rangle \langle \psi|
M_m^{\dagger} \right).
\end{eqnarray}
Since $|\phi\rangle$ is a pure state, it follows that
\begin{eqnarray}
{\cal E}^B_m \left( M_m |\psi\rangle \langle \psi| M_m^{\dagger} \right) \propto
|\phi\rangle \langle \phi|.
\end{eqnarray}
Tracing out system $B$ gives $M_m \rho_\psi M_m^{\dagger} \propto
\rho_\phi$, with non-negative constants of proportionality $p_m$
satisfying $\sum_m p_m = 1$.  Polar decomposing $M_m
\sqrt{\rho_{\psi}}$ gives:
\begin{eqnarray}  \label{eqtn:inter_1}
M_m \sqrt{\rho_\psi} = \sqrt{ M_m \rho_\psi M_m^{\dagger}} U_m =
\sqrt{p_m} \sqrt{\rho_\phi} U_m, 
\end{eqnarray}
where $U_m$ is a unitary matrix.  But $\sum_m M_m^{\dagger} M_m = I$,
from which we obtain:
\begin{eqnarray} \label{eqtn:inter_2}
\rho_\psi = \sum_m \sqrt{\rho_\psi} M_m^{\dagger} M_m \sqrt{\rho_\psi}.
\end{eqnarray}
Substituting equation~(\ref{eqtn:inter_1}) and its adjoint into
equation~(\ref{eqtn:inter_2}) gives $\rho_\psi = \sum_m p_m
U_m^{\dagger} \rho_\phi U_m$, and fact~\ref{fact:random_unitary}
implies that $\lambda_\psi \prec \lambda_\phi$, as required.

%
%
To prove the converse, we consider first the two dimensional case,
which demonstrates the essential idea of the general proof.  Using
orthonormal basis states $|0\rangle$ and $|1\rangle$, and the Schmidt
decomposition, we may always write
\begin{eqnarray} \label{eqtn:converse_1} 
|\psi\rangle \sim |\psi'\rangle = \sqrt{\alpha_+} |00\rangle + \sqrt{\alpha_-} |11\rangle. 
\end{eqnarray} 
where $0 \leq \alpha_- \leq \alpha_+ \leq 1$, and $\alpha_+ + \alpha_-
= 1$.  Since $\lambda_\psi \prec \lambda_\phi$, we may choose
non-negative $\beta_{\pm}$ which sum to one, such that $\beta_- \leq
\alpha_-$ and $\alpha_+ \leq \beta_+$, and 
\begin{eqnarray} \label{eqtn:converse_2} 
|\phi\rangle \sim |\phi'\rangle = \sqrt{\beta_+} |00\rangle + \sqrt{\beta_-}|11\rangle.
\end{eqnarray} 
The first step of the protocol is to transform $|\psi\rangle$ to
$|\psi'\rangle$, which Alice and Bob may do with local unitary
operations.  A simple eigenvalue calculation (fact~\ref{fact:Schmidt}) shows
that
\begin{eqnarray}
|\psi'\rangle \sim |\psi''\rangle \equiv
\frac{|00\rangle+|1\rangle(\cos(\gamma)|0\rangle+\sin(\gamma)|1\rangle)}{\sqrt 2},
\end{eqnarray}
where $\gamma$ is chosen to satisfy $\alpha_+ = (1+\cos(\gamma))/2$.
The next step of the protocol is for Alice and Bob to transform
$|\psi'\rangle$ to $|\psi''\rangle$, again by local unitary operations on
their respective systems.  Next, define operators $M_1$ and $M_2$ on
Alice's system to have the following matrix representations in the
$|0\rangle, |1\rangle$ basis:
\begin{eqnarray}
M_1  =  \left[ \begin{array}{cc} 
\cos(\delta) & 0 \\ 0 & \sin(\delta) \end{array} \right]; \,\,\,\,
M_2  =  \left[ \begin{array}{cc} 
\sin(\delta) & 0 \\ 0 & \cos(\delta) \end{array} \right].
\end{eqnarray}
$\delta$ is a parameter whose exact value will be fixed later in the
proof.  Note that $M_1^{\dagger} M_1 + M_2^{\dagger}M_2 = I$, so this
defines a generalized measurement on Alice's system, which may be
implemented using standard techniques involving only projective
measurements and unitary transforms \cite{Schumacher96a}.  Let
$|\psi'''_m\rangle$ denote the state after the measurement, given that
outcome $m$ occurred.  Then
\begin{eqnarray}
|\psi'''_1\rangle & = &
\cos(\delta)|00\rangle+\sin(\delta)|1\rangle \left( \cos(\gamma)|0\rangle+
\sin(\gamma)|1\rangle \right) \nonumber \\
& & \\
|\psi'''_2\rangle & = &
\sin(\delta)|00\rangle+\cos(\delta)|1\rangle \left( \cos(\gamma)|0\rangle+
\sin(\gamma)|1\rangle \right). \nonumber \\
& & 
\end{eqnarray}
By symmetry or explicit eigenvalue calculation, one may verify that
$|\psi'''_1\rangle \sim |\psi'''_2\rangle$.  Thus Alice and Bob can ensure
that the final state is $|\psi'''\rangle \equiv |\psi'''_1\rangle$, by
applying appropriate unitary transforms to their respective systems.
To do this, Alice must send her measurement result to Bob, so he knows
which unitary operation to apply.  An eigenvalue calculation
shows that
\begin{eqnarray}
|\psi'''\rangle \sim \sqrt{\lambda_+}|00\rangle +\sqrt{\lambda_-}|11\rangle, \end{eqnarray}
where
\begin{eqnarray}
\lambda_{\pm} \equiv
\frac{1\pm\sqrt{1-\sin^2(2\delta)\sin^2(\gamma)}}{2}.
\end{eqnarray}
At $\delta = 0$, $\lambda_+ = 1$ and at $\delta = \pi/4$, $\lambda_+ =
(1+\cos(\gamma))/2 = \alpha_+$.  Since $\alpha_+ \leq \beta_+ \leq 1$,
continuity ensures that the equation $\lambda_+(\delta) = \beta_+$ has
a real solution $\delta = \frac 12
\arcsin\left(2(\beta_+-\beta_+^2)^{1/2}/\sin \gamma\right)$.
Choosing this $\delta$ gives $|\psi'''\rangle \sim |\phi'\rangle =
\sqrt{\beta_+}|00\rangle+\sqrt{\beta_-}|11\rangle$, and therefore by applying
local unitary transformations Alice and Bob may obtain the state
$|\phi'\rangle$, and from there the state $|\phi\rangle$, by
equation~(\ref{eqtn:converse_2}).

%
%
The general case uses fact~\ref{fact:t-transforms} to reduce the
problem to the two dimensional case by cascading a sequence of
entanglement transformations, each corresponding to a single
T-transform.  Using facts~\ref{fact:t-transforms}
and~\ref{fact:Schmidt} we may assume that Alice and Bob are each in
possession of a $d$-dimensional system, with orthonormal bases
$|0\rangle,|1\rangle,\ldots,|d-1\rangle$, that the state
$|\psi\rangle$ has the form
\begin{eqnarray}
|\psi\rangle \sim |\psi'\rangle & = & \cos(\zeta) \left( \sqrt{\alpha_+} |00\rangle +
\sqrt{\alpha_-} |11\rangle \right) + \sin(\zeta) |\psi_{\perp}\rangle,
\nonumber \\
& &
\end{eqnarray} 
where $|\psi_{\perp}\rangle$ is a normalized state of the form
$\sum_{j=2}^{d-1} \psi_j |j\rangle|j\rangle$, $\zeta$ is real, and
\begin{eqnarray}
|\phi\rangle \sim |\phi'\rangle & = & \cos(\zeta) \left( \sqrt{\beta_+} |00\rangle + 
\sqrt{\beta_-}|11\rangle \right) + \sin(\zeta) |\psi_{\perp}\rangle. \nonumber
\\
& & 
\end{eqnarray} 
$0 \leq \beta_- \leq \alpha_- \leq \alpha_+ \leq \beta_+$, as before.
The T-transform corresponds to a transformation of the $|00\rangle$
and $|11\rangle$ terms in these expressions.  The protocol is as for
the two-dimensional case, except for a slight change at the
measurement stage. Alice does a generalized measurement described by
operators $\tilde M_1$ and $\tilde M_2$ defined in terms of the
earlier operators $M_1$ and $M_2$ by
\begin{eqnarray}
\tilde M_1  =  \left[ \begin{array}{cc} M_1 & 0 \\ 0 &
    \frac{I_{d-2}}{\sqrt 2}
\end{array} \right]; \,\,\,\,
\tilde M_2 =  \left[ \begin{array}{cc} M_2 & 0 \\ 0 &
    \frac{I_{d-2}}{\sqrt 2}
\end{array} \right].
\end{eqnarray}
The matrices $I_{d-2}/\sqrt 2$ in the lower right hand corner ensure
that coherence is preserved during the transformation procedure, and
the completeness relation $\sum_m \tilde M_m^{\dagger}
\tilde M_m = I$ is obeyed.  With this change the protocol
proceeds as before to transform $|\psi\rangle$ to $|\phi\rangle$.

%
%
The next few paragraphs examine some consequences of Theorem~1.  Note
first that the proof of the Theorem, together with the method given in
\cite{Bhatia97a} for obtaining $\lambda_\psi$ from T-transforms acting on
$\lambda_\phi$, gives a constructive method involving at most $d-1$
bits of communication to transform $|\psi\rangle$ to $|\phi\rangle$, whenever
$\lambda_\psi \prec \lambda_\phi$.

%
%
Generalizing the earlier example of incomparable states $|\psi\rangle$
and $|\phi\rangle$, I conjecture that in the limit where $A$ and $B$
are of large dimensionality, almost all pairs of pure states
$|\psi\rangle$ and $|\phi\rangle$ picked according to the unitary
invariant measure on $AB$ \cite{Page93a} will be incomparable.  A
heuristic argument is as follows.  Let $p_i$ and $q_i$ be random
variables denoting the eigenvalues of $\rho_\psi$ and $\rho_\phi$,
arranged into decreasing order.  Define $\Delta_i \equiv p_i-q_i$.
Then $|\psi\rangle$ and $|\phi\rangle$ are incomparable if (and only
if) the stochastic process $T_k \equiv
\sum_{i=1}^k \Delta_i$ {\em crosses the origin}, that is, it is
positive for some values of $k$, and negative for others.  If $T_k$
were a random walk with independent and identically distributed
increments, the conjecture would be true in the limit of large
dimension \cite{Grimmett92a}.  $T_k$ fails to be a random walk for two
reasons: (a) The ordering of the $p_i$ and $q_i$ ensures that the
typical size of the increments $\Delta_i$ tends to decrease as $i$
gets larger; and (b) The constraint $\sum_{i=1}^d
\Delta_i = 0$ ensures that the increments are correlated.
Intuitively, in the limit of large dimensionality, the distribution of
the $\Delta_i$ becomes very nearly uncorrelated from step to step,
with the remaining correlations acting as a weak ``restoring force''
towards the origin, which tends to enhance crossings.  Furthermore,
the distribution ``flattens out'' in large dimensions, with only a
very slow decrease in the typical size of the increments
\cite{Lubkin78a}.  So in large dimensions the $\Delta_i$ behave
locally like increments of a random walk, which can therefore be
expected to cross the origin.

%
%
Theorem 1 allows the well-developed theory of {\em isotone functions}
\cite{Bhatia97a} to be applied to the study of entanglement.  For
example, an important subclass of the isotone functions is the {\em
Schur-convex functions}: $f : R^d \rightarrow R$ such that $x \prec y$
implies $f(x) \leq f(y)$.  Well-known Schur-convex functions
\cite{Bhatia97a} include the maps $\{ x_i \} \rightarrow
\sum_i x_i \log x_i$ and $\{ x_i \}
\rightarrow \sum_i x_i^k$, for any $k \geq 1$.
It follows that if $|\psi\rangle \rightarrow |\phi\rangle$, then $S(\rho_\phi)
\leq S(\rho_\psi)$, where $S(\cdot)$ is the von Neumann entropy,
and $\mbox{tr}(\rho_\psi^k) \leq \mbox{tr}(\rho_\phi^k)$, for any $k \geq 1$.

%
%
Theorem~1 simplifies in the special case where Alice's system is two
dimensional, and Bob's system arbitrary, telling us that $|\psi\rangle
\rightarrow |\phi\rangle$ if and only if $S(\rho_\phi) \leq
S(\rho_\psi)$.

%
%
Theorem 1 may be combined with the asymptotic equipartition theorem
(Chapter~3 of \cite{Cover91a}) to provide a straightforward proof of
some results of Bennett {\em et al} \cite{Bennett96c}.  They showed
how to approximately transform back and forth between $n$ copies of
the state $|\phi\rangle$ and $nS(\rho_\phi)$ EPR pairs, in the limit
where $n$ becomes large.  The following is a sketch of the proof based
upon Theorem~1.

%
%
Suppose Alice and Bob share $m$ EPR pairs.  Denote their total state
by $|\psi\rangle$, which has corresponding vector of eigenvalues
$(2^{-m},2^{-m},\ldots,2^{-m})$.  Let $|\phi\rangle$ be any pure state
of $AB$.  Taking $n$ copies of $|\phi\rangle$, the asymptotic
equipartition theorem implies that for sufficiently large $n$ the
state may be approximated by just $2^{nS(\rho_\phi)}$ terms in the
Schmidt decomposition,
\begin{eqnarray}
|\phi\rangle^{\otimes n} \approx |\phi'\rangle \equiv
\sum_{i=1}^{2^{nS(\rho_\phi)}} \sqrt{\lambda_i} |i\rangle |i\rangle.
\end{eqnarray}
Choose $m$ such that $m \approx nS(\rho_\phi)$.  Then it is easy to
check directly that $(2^{-m},\ldots,2^{-m}) \prec
(\lambda_1,\ldots,\lambda_{2^{nS(\rho_\phi)}})$, so Theorem~1 implies
that $|\psi\rangle \rightarrow |\phi'\rangle$, and thus it is possible
to transform $n S(\rho_{\phi})$ EPR pairs into a pretty good
approximation to $n$ copies of $|\phi\rangle$.

%
%
For the inverse procedure note that by the asymptotic equipartition
theorem there is a set of roughly $2^{nS(\rho_\phi)}$ terms in the
Schmidt decomposition such that
\begin{eqnarray}
|\phi\rangle^{\otimes n} \approx \sum\nolimits^{\prime}
\sqrt{\lambda_i} |i\rangle|i\rangle, 
\end{eqnarray}
where the primed sum indicates that we are summing over a restricted
set where $\lambda_i {\scriptstyle{\stackrel{<}{\sim}}}
2^{-nS(\rho_\phi)}$.  The transformation procedure is for Alice to
first project onto the space spanned by the terms $|i\rangle$
appearing in the sum.  This succeeds with probability $1-\epsilon
\approx 1$, leaving the state in the form
\begin{eqnarray}
\sum\nolimits^{\prime} \sqrt{\frac{\lambda_i}{1-\epsilon}} |i\rangle|i\rangle
\end{eqnarray}
Therefore, for any $m$ such that $2^{-nS(\rho_\phi)}/(1-\epsilon)
\leq 2^{-m}$ Theorem~1 implies that the $n$ copies of $|\phi\rangle$
may be transformed to $m$ EPR pairs.  In particular, we may choose $m
\approx nS(\rho_\phi)$.

%
%
There are many open problems to which Theorem~1 may be of relevance.
It would be of great interest to determine when a mixed state $\rho$
can be transformed to a mixed state $\sigma$ by local operations and
classical communication.  This would also provide a good starting
point to better understand {\em approximate} entanglement
transformation.  A related problem is to determine transformation
conditions for three (or more) party pure state entanglement analogous
to those found here for two party entanglment.  Finally, I hope that
the connection between entanglement and majorization may enable us to
better understand the fundamental measures of entanglement introduced
by Wootters and collaborators
\cite{Wootters98a,Bennett96a}.

\section*{acknowledgments}
Thanks to Howard Barnum, who introduced me to the beautiful subject of
majorization, to Dorit Aharonov and Bill Wootters, whose words and
papers convinced me that entanglement for its own sake is a deeply
interesting subject, and to Ike Chuang, Chris Fuchs, Julia Kempe, and
John Preskill for helpful discussions.  This work was supported by a
Tolman Fellowship, and by DARPA through the Quantum Information and
Computing Institute (QUIC) administered through the ARO.

\end{multicols}

\end{document}